\documentclass[aps,prd,twocolumn,nofootinbib,superscriptaddress,preprintnumbers,balancelastpage,longbibliography]{revtex4-1}

\usepackage[utf8]{inputenc}              
\usepackage[english]{babel}              
\usepackage{amsmath, amsfonts, mathtools, physics, slashed, xfrac}
\usepackage[normalem]{ulem}  
\usepackage{graphicx}
\usepackage{graphics}                     
\usepackage{tabularx}
\usepackage{booktabs}
\usepackage{multirow}
\usepackage{makecell}
\usepackage{dcolumn}
\usepackage{siunitx}                    
\usepackage{fancyhdr}
\usepackage{hyperref}
\usepackage{lettrine}
\usepackage[dvipsnames]{xcolor}

\input Zallman.fd

\hypersetup{
     colorlinks   = true,
     citecolor    = darkviolet,
     urlcolor     = darkviolet,
     linkcolor    = darkviolet
}

\definecolor{darkblue}{rgb}{0.0,0.0,0.75}
\definecolor{darkred}{rgb}{0.6,0.0,0}
\definecolor{darkgreen}{rgb}{0.0,0.6,0.}
\definecolor{darkviolet}{rgb}{0.58, 0.0, 0.83}

\newcommand\redsout{\bgroup\markoverwith{\textcolor{red}{\rule[0.5ex]{2pt}{0.4pt}}}\ULon}

\begin{document}

\preprint{SLAC-PUB-251210}

\title{Sub-GeV Dark Matter Detection with Dark Rates in Liquid Scintillators}

\author{Lillian Santos-Olmsted}
\thanks{\href{mailto:solmsted@stanford.edu}{solmsted@stanford.edu}, \href{https://orcid.org/0000-0002-8763-3702}{0000-0002-8763-3702}}
\affiliation{Particle Theory Group, SLAC National Accelerator Laboratory, Stanford, CA 94305, USA}
\affiliation{Kavli Institute for Particle Astrophysics and Cosmology, Stanford University, Stanford, CA 94305, USA}

\author{Rebecca~K.~Leane}
\thanks{\href{mailto:rleane@slac.stanford.edu}{rleane@slac.stanford.edu}, \href{https://orcid.org/0000-0002-1287-8780}{0000-0002-1287-8780}}
\affiliation{Particle Theory Group, SLAC National Accelerator Laboratory, Stanford, CA 94305, USA}
\affiliation{Kavli Institute for Particle Astrophysics and Cosmology, Stanford University, Stanford, CA 94305, USA}

\author{Carlos Blanco}
\thanks{\href{mailto:carlosblanco2718@princeton.edu}{carlosblanco2718@princeton.edu}, \href{https://orcid.org/0000-0001-8971-834X}{0000-0001-8971-834X}}
\affiliation{Institute for Gravitation and the Cosmos, The Pennsylvania State University, University Park, PA 16802, USA}
\affiliation{Department of Physics, Princeton University, Princeton, NJ 08544, USA}
\affiliation{Stockholm University and The Oskar Klein Centre for Cosmoparticle Physics, Alba Nova, 10691 Stockholm, Sweden}

\author{John F. Beacom}
\thanks{\href{mailto:beacom.7@osu.edu}{beacom.7@osu.edu}; \href{https://orcid.org/0000-0002-0005-2631}{0000-0002-0005-2631}
}
\affiliation{Center for Cosmology and AstroParticle Physics (CCAPP),
Ohio State University, Columbus, Ohio 43210, USA}
\affiliation{Department of Physics, Ohio State University, Columbus, Ohio 43210, USA}
\affiliation{Department of Astronomy, Ohio State University, Columbus, Ohio 43210, USA}

\date{\today}

\begin{abstract}
It was recently shown that standard sub-GeV dark matter candidates can be effectively probed by large neutrino observatories via annual modulation of the total photomultiplier hit rate. That work focused on the production of light by the excitation of scintillator molecules and considered the JUNO detector, surpassing limits from dedicated dark-matter detectors and reaching theoretical targets. Here, we significantly generalize that work, now also taking into account ionization channels and extending the analysis to other liquid-scintillator detectors, including SNO+, Daya Bay, Borexino, and KamLAND. Last, we present a call to action: with multiple detectors achieving competitive sensitivity, there is an opportunity to validate this new technique across experiments and to refine it using each detector’s strengths.

\end{abstract}

\maketitle


\section{Introduction}

\lettrine{S}{ub-gev dark-matter (dm) candidates} are attracting strong interest, driven by compelling theoretical motivations and wide swaths of unexplored parameter space~\cite{Alfonso-Pita:2022akn, Krnjaic:2022ozp, SuperCDMS:2022kse, Akesson:2022vza, Mitridate:2022tnv, Wang:2022cyk, Battaglieri:2017aum}. The main obstacle for direct detection at low masses is that nuclear recoil energies become too small for current technologies. This has shifted attention to other avenues, such as DM-electron scattering, which features more favorable kinematics, as well as a broad range of new experiments, including novel uses of superconductors, superfluids, organic targets, quantum devices, and more~\cite{Essig:2011nj, Graham:2012su, Essig:2015cda, Hochberg:2015pha, Hochberg:2015fth, Hochberg:2019cyy, Derenzo:2016fse, Schutz:2016tid, Hochberg:2016ntt, Essig:2016crl, Hochberg:2017wce, Cavoto:2017otc, Emken:2017erx, Emken:2017qmp, Griffin:2018bjn, Sanchez-Martinez:2019bac, Essig:2019xkx, Emken:2019tni,  Kurinsky:2019pgb, Geilhufe:2019ndy, Blanco:2019lrf, Baxter:2019pnz, Catena:2019gfa, Radick:2020qip,Gelmini:2020xir, Trickle:2020oki, Kurinsky:2020dpb, Griffin:2020lgd, Blanco:2021hlm, Knapen:2021run, Hochberg:2021ymx, Hochberg:2021yud, Essig:2022dfa, Hochberg:2022apz,Das:2022srn, Das:2024jdz, Griffin:2024cew,Cook:2024cgm, Simchony:2024kcn,  QROCODILE:2024zmg, Marocco:2025eqw, Blanco:2022cel}. Despite these advances, achieving the large exposures needed for a decisive discovery remains an open challenge.

In principle, one way to increase exposure is to take advantage of the huge volumes of neutrino detectors, some of which are on the kiloton scale, as opposed to the ton scale of DM detectors. However, neutrino detectors are designed to look for events that reach typical deposited energies on the order of MeV and produce a specific hit pattern on many photomultiplier tubes (PMTs). For standard DM scattering, nuclear recoils are at the keV scale and electron recoils are at the eV scale. Therefore, to overcome thresholds, DM searches using neutrino detectors have focused on specific cases such as heavy or boosted DM~\cite{Agashe:2015xkj, Kopp:2015bfa, Ema:2018bih, Bringmann:2018cvk, Cappiello:2019qsw, Wang:2021jic, Toma:2021vlw, PROSPECT:2021awi,  Aoki:2023tlb, Bell:2023sdq, Dutta:2024kuj, Jeesun:2025gzt, BetancourtKamenetskaia:2025noa, Diurba:2025lky,DeMarchi:2025uoo, Bramante:2018tos, Eby:2019mgs, Bramante:2019yss, Ponton:2019hux, Church:2020env, Bai:2022nsv, Aggarwal:2024ngx}.

\begin{figure}[t]
    \centering
    \includegraphics[width=0.9\columnwidth]{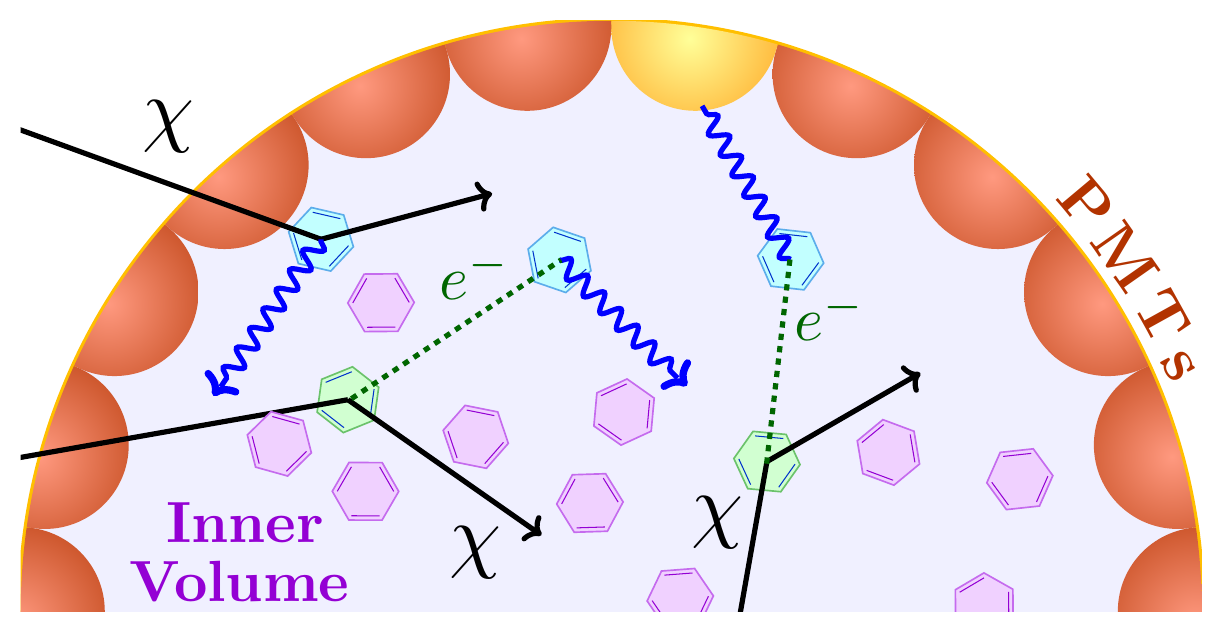}
    \caption{Diagram showing DM-induced excitation (prior work \cite{Leane:2025efj}) and ionization (this work) in organic scintillator molecules, leading to detection by PMTs in a neutrino observatory.}
    \label{fig:detector}
\end{figure}

That changed with Ref.~\cite{Leane:2025efj}, which pointed out that neutrino observatories can be repurposed as DM direct detection experiments even for standard DM scenarios. This can be achieved by taking advantage of DM directly exciting target molecules in the observatory and producing enough fluorescence light \textit{in aggregate over many DM scatterings} to be detected over the inherent noise of the PMTs. Because this approach does not rely on MeV-scale events or a pattern of PMT hits, the main factors that determine the detection of a DM signal are the excitation thresholds of the target, and the production and propagation of light in the detector medium. The organic liquid scintillators used by many neutrino detectors have lower excitation thresholds than the noble liquids used by many dedicated DM detectors, making them optimal in this regard~\cite{Leane:2025efj}. Although the noise of the PMTs in neutrino detectors is large relative to the DM signal, this challenge can be reduced by using the annual modulation of the DM velocity relative to Earth. In Ref.~\cite{Leane:2025efj}, these techniques were used to compute the projected sensitivity to the DM-electron cross section from the fluorescence of bound-state molecular transitions in neutrino detectors, with an emphasis on the Jiangmen Underground Neutrino Observatory (JUNO). 

In this paper, we improve the fluorescence treatment, and we additionally calculate the contribution of ionization to the DM signal rate, which adds sensitivity especially for larger DM masses in the case of a heavy mediator. We also extend the combined analysis to other liquid scintillators, namely, SNO+, Daya Bay, Borexino, and KamLAND. We show that all five neutrino detectors have competitive sensitivity to sub-GeV DM. The value in this lies not only in improving the sensitivity, but also in the opportunity for multiple experiments to verify each other's results (with differing site-/detector-specific systematics) and for each detector to independently refine this new technique.

This paper is organized as follows. In Sec.~\ref{sec:detectors}, we review the liquid scintillators mentioned above. In Sec.~\ref{sec:sigRate}, we calculate the signal rate from excitation and ionization, including a discussion of detection efficiency. In Sec.~\ref{sec:sensitivity}, we present the resulting sensitivity. We conclude in Sec.~\ref{sec:conc} with a summary of our results and a discussion of the importance of improving and implementing this technique.


\section{Review of Neutrino detectors}
\label{sec:detectors}

We now discuss each of the neutrino detectors we consider. We focus on liquid-scintillator based detectors due to their high light yield and detection efficiency. Table~\ref{tab:detector_comparison} summarizes the primary features of each detector.

\begin{table*}[ht]
\centering
{\fontsize{9pt}{13pt}\selectfont
\renewcommand{\arraystretch}{1.2}
\setlength{\tabcolsep}{10pt}
\begin{tabular}{|l|c|c|c|c|c|c|}
\hline
\textbf{Property} &
\textbf{JUNO} &
\textbf{SNO+} &
\textbf{Daya Bay} &
\textbf{Borexino} &
\multicolumn{2}{c|}{\textbf{KamLAND}} \\
\hline
\makecell[l]{Main target} & LAB & LAB & LAB & PC & PC & Dodecane \\
\hline
\makecell[l]{Target mass [tons]} & 20,000 & 780 & 320 & 278 & 200 & 800 \\
\hline
\makecell[l]{Number of molecules} & $5 \times 10^{31}$ & $2 \times 10^{30}$ & $8 \times 10^{29}$ & $1.3 \times 10^{30}$ & $9.1 \times 10^{29}$ & $2.6 \times 10^{30}$ \\
\hline
\makecell[l]{Dark rate [Hz]} & $6 \times 10^8$ & $10^7$ & $2.3 \times 10^6$ & $8.8 \times 10^5$ & \multicolumn{2}{c|}{$1.3 \times 10^7$} \\
\hline
\makecell[l]{PMT coverage ($X$)} & 0.78 & 0.54 & 0.12 & 0.34 & \multicolumn{2}{c|}{0.34} \\
\hline
\makecell[l]{Efficiency ($Y$)} & 0.28 & 0.14 & 0.22 & 0.30 & \multicolumn{2}{c|}{0.20} \\
\hline
\makecell[l]{Light yield ($LY$) [$\gamma/\text{keV}$]} & 10 & 11.9 & 10 & 11.5 & \multicolumn{2}{c|}{8} \\
\hline
\makecell[l]{Photoelectron yield (PE) [$e^{-}/\text{keV}$]} & 1.7 & 0.3 & 0.165 & 0.5 & \multicolumn{2}{c|}{0.3} \\
\hline
\makecell[l]{Runtime (years)} & 10 & 3 & 9 & 14 & \multicolumn{2}{c|}{23} \\
\hline
\end{tabular}
}
\caption{Comparison of neutrino detector properties. The primary target in each detector is linear alkylbenzene (LAB), pseudocumene (PC), or dodecane. For KamLAND, the target mass and number of molecules are split by the PC and dodecane components. The efficiencies, light yields, and photoelectron yields are discussed in Sec.~\ref{sec:LY}. For JUNO, we assume a nominal runtime of 10 years.  For the other detectors, we use the data collection periods to date.}
\label{tab:detector_comparison}
\end{table*}


\subsection{JUNO}

JUNO~\cite{JUNO:2021vlw, JUNO:2020xtj} is located 700~m underground in southern China and began collecting data this year~\cite{JUNO:2025fpc, JUNO:2025gmd}. In our results for JUNO, we assume a nominal runtime of 10 years. JUNO is well-equipped to achieve its goals of studying the neutrino mass ordering and other fundamental neutrino properties. Its detector consists of a 35-m diameter spherical acrylic vessel containing 20,000 tons of target material. 

The primary target is linear alkyl benzene (LAB). Each LAB molecule is made up of carbon and hydrogen atoms arranged in a benzene ring with an alkyl chain. Its chemical formula is $ \text{C}_6\text{H}_5\text{C}_n\text{H}_{2n+1}$, where $n$ is $10-13$ for JUNO~\cite{Lombardi:2019epz}. To be conservative, we set $n=10$ when determining the number of atoms in each LAB molecule. In addition to LAB, there are small portions of other substances: 2.5~g/L of 2,5-diphenyloxazole (PPO) as the primary fluor and 3~mg/L of p-bis-(o-methylstyryl)-benzene (bis-MSB) as the secondary wavelength shifter~\cite{JUNO:2020bcl,Beretta:2025rwj}. The target material is surrounded by 17,612 20-inch PMTs and 25,600 3-inch PMTs, with a total effective wall coverage of 78\%~\cite{JUNO:2021vlw}. The total dark rate in JUNO is roughly $6 \times 10^8$~Hz~\cite{JUNO:2022hlz, Zhang:2023dha, JUNO:2025dfn}. 


\subsection{SNO+}

As the successor to the Sudbury Neutrino Observatory (SNO)~\cite{SNO:1999crp}, SNO+~\cite{SNO:2020fhu, SNO:2021xpa} is located 2 km underground at the SNOLAB facility in Sudbury, Canada and uses much of the SNO infrastructure. SNO+ has been in its liquid scintillator phase since 2022~\cite{Inacio:2024yfm}. Its main purpose is to search for neutrinoless double beta decay, but it is also sensitive to geoneutrinos, reactor antineutrinos, solar neutrinos, and supernova neutrinos. The main component of the detector is a 6-m radius acrylic vessel filled with target material, which is 780 tons of LAB doped with 2.2~g/L of PPO. In late 2023, 2.2~mg/L of bis-MSB was also added to the scintillator~\cite{SNO:2025chx}. The target volume is surrounded by 9,362 8-inch PMTs, with an effective wall coverage of 54\% \cite{Inacio:2024yfm}. Each PMT has a dark rate of about 1~kHz, leading to a total dark rate of $\sim$$10^7$~Hz~\cite{Marzec:2019iwu}. In the upcoming tellurium phase, a small concentration of natural tellurium ($0.5 \%$ by mass) will be added to the scintillator \cite{Inacio:2024yfm}. Note that this will decrease the light yield by a few tens of percent~\cite{Auty:2022lgh}.  


\subsection{Daya Bay}

The Daya Bay reactor neutrino experiment detectors ~\cite{Band:2013zka, Band:2012dh, Cao:2016vwh, Li:2024zzv} are located in Shenzhen, China, with depths ranging from 250--860 meters water equivalent. During its runtime between 2011 and 2020, Day Bay's primary goal was to measure the neutrino oscillation parameter $\sin^2 (2\theta_{13})$ at high precision from inverse beta decay events. Daya Bay detects antineutrinos from six nuclear reactors, and is made up of four pairs of underground antineutrino detectors distributed across three experimental halls, located between $300$~m and $2$~km from the nuclear reactors. 

Each of the eight antineutrino detectors is cylindrical and has three zones nested within each other. The outer 5-m radius zone is a stainless steel vessel that houses the PMTs and is filled with mineral oil to protect the inner zones from radiation. The middle zone is an acrylic vessel with a radius of $4$~m, filled with LAB. Finally, the innermost zone (the fiducial volume for reactor studies) is a 3-m acrylic vessel containing LAB doped with gadolinium. Because our search is sensitive to the total PMT rates, our target volume includes both the innermost and middle zones, providing a total of 320 tons of target material among the eight detectors. There are also 3~g/L of PPO and 15~mg/L of bis-MSB in the target medium as primary and secondary fluors, respectively~\cite{Beriguete:2014gua}. Daya Bay has 1,536 PMTs, which provide an effective wall coverage of 12\%~\cite{Tsang:2013tcv}. Each PMT has a dark rate of $1.5$~kHz~\cite{DayaBay:2015meu}, producing a total dark rate of $\sim$$2.3 \times 10^6$~Hz.


\subsection{Borexino}

Borexino~\cite{Borexino:2013zhu,BOREXINO:2023ygs} is located in the Laboratori Nazionali del Gran Sasso in Italy, at a depth of 3800~meters water equivalent. It collected data from 2007 to 2021, and its physics objectives focused on the detection of low-energy solar neutrinos. In Borexino, the target medium is composed of pseudocumene (PC, 1,2,4-trimethylbenzene) doped with 1.5~g/L of PPO. Like LAB, PC is a benzene-based aromatic hydrocarbon, and its chemical formula is $\text{C}_9 \text{H}_{12}$. The 278-ton target material is contained in a spherical 4.25-m nylon vessel, surrounded by 2212 internal PMTs with a wall coverage of 34\%~\cite{Leung:2006kk}. Each PMT has a dark rate of about $400$~Hz, resulting in a total dark rate of $8.8 \times 10^5$~Hz~\cite{Borexino:2013zhu}. 


\subsection{KamLAND} 

The Kamioka Liquid-scintillator AntiNeutrino Detector (KamLAND)~\cite{Suekane:2004ny, Suzuki:2014woa, KamLAND:2021gvi} is located 1~km underground in Kamioka, Japan. KamLAND has been collecting data since 2002, with the goal of studying low-energy neutrino oscillations by detecting reactor antineutrinos. The KamLAND detector consists of a 13-m diameter sphere made of nylon and ethylene vinyl alcohol, filled with one kiloton of the target material. In KamLAND, the target medium is 80\% dodecane (a linear alkane with the chemical formula $\text{C}_{12}\text{H}_{26}$) and 20\% PC, doped with 1.36~g/L of PPO. 

In 2011, the KamLAND detector was updated to incorporate the KamLAND-Zen double-beta decay experiment \cite{Gando:2012psd}. A 1.5-m radius balloon, holding 13 tons of xenon-loaded liquid scintillator, was inserted into the detector. However, KamLAND continues to run normally \cite{Decowski:2016axc, Eizuka:2024rhh}. For our purposes, it is not necessary to include the xenon component of KamLAND-Zen in our analysis. Although the liquid scintillator in the balloon has a similar ratio of dodecane to PC as the main KamLAND detector, and this material could in principle be incorporated into the target volume, there would only be a $\sim$$1 \%$ gain. Moreover, the xenon itself would not contribute to increasing the light yield or improving detector efficiency. 

The spherical vessel containing the target material is enclosed by mineral oil to protect the inner components from radiation. The oil is surrounded by 1,325 17-inch PMTs and 554 20-inch PMTs, with a total PMT wall coverage of 34\%~\cite{Suekane:2004ny}. Each PMT has a dark rate of 7~kHz~\cite{Kozlov:2006zz}, giving a total dark rate of $1.3 \times 10^7$~Hz. 


\section{Dark Matter Signal Rate Calculation}
\label{sec:sigRate}

We now calculate the signal rate of PMT hits originating from DM-induced excitation and ionization events. 


\subsection{Total Event Rate}

Liquid scintillator detectors are sensitive to particle energy depositions in two distinct ways. First, if the recoil energy of the DM scattering is less than the ionization or dissociation threshold, the molecule can undergo an internal electronic excitation, producing a photon emission signal in the form of fluorescence. Second, more energetic DM particles may also ionize the molecule. The relaxation of many ionized molecules or atoms can lead to a burst of light that generates a signal in the form of scintillation.

The total event rate is given by the sum of the excitation and ionization contributions,
\begin{equation}
\label{eq:totRate}
    R_{\rm{tot}} = R_{\rm{exc}}+R_{\rm{ion}},
\end{equation}
where $R_{\rm{exc}}$ is the event rate due to transitions between ground and excited electronic states, and $R_{\rm{ion}}$ is the event rate from ionization. 


\subsection{Excitation Rate}
\label{sec:update}

Following Ref.~\cite{Leane:2025efj}, we calculate the total event rate from excitation for spin-independent DM-SM interactions as 
\begin{equation}
\label{eq:excRate}
    R_{\text{exc}} = \xi_{\text{exc}}\frac{N_T^{\text{exc}}\rho_\chi \bar{\sigma}_e}{8\pi m_{\chi}\mu_{\chi e}^2}\sum_{i,j}\int \frac{d^3q}{q}\eta(v_{\text{min}}^{ij})F_{\text{DM}}^2(q)|f_{ij}(\vec{q})|^2,
\end{equation}
where $m_{\chi}$ is the DM mass, $\rho_{\chi} = 0.4 \ \rm{GeV}/cm^3$ is the local DM density, $\mu_{e \chi}$ is the DM-electron reduced mass, $\bar{\sigma}_e$ is the DM-electron cross section, $N_T^{\text{exc}}$ is the number of targets, and $\xi_{\rm{exc}}$ is the efficiency factor. In the summation, the indices correspond to the transitions between the bound states $i$ and $j$. The three functions in the momentum transfer $q$ integral are the excitation form factor $f_{ij}(\vec{q})$, the DM form factor $F_{\rm DM}$, and the mean inverse velocity of the DM $\eta(v_{\rm min}^{ij}(q))$. We now detail each of these terms separately.

The material properties are encapsulated by the form factors $f_{ij}(\vec{q})$. For LAB and PC, we use the form factor of benzene, because the transitions that occur for electrons in the benzene ring are expected to dominate~\cite{Leane:2025efj, Blanco:2019lrf}. We use the excitation form factor for benzene from Ref.~\cite{Blanco:2019lrf}, which was developed to agree with UV spectroscopy. The molecular orbital model is based on semi-empirical post-Hartree-Fock calculations. The electron-electron interaction is taken into account via configurational interactions, leading to states of known symmetry that are described by combinations of Slater determinants~\cite{pariser1953semi,pariser1953semi2}. The energies of the transitions are adjusted to fit the experimentally determined UV absorption bands of the benzene-based scintillator~\cite{katz1971electronic}. 

We adopt a conservative accounting of target electrons in the case of excitation (fluorescence). Because the primary character of the fluorescence in substituted benzenes comes from $\pi\rightarrow\pi^*$ transitions in the aromatic ring, we do not include electrons that are either in the core shells of the atoms or in $\rm{sp}^3$ hybridized orbitals. In other words, only electrons that participate in the aromatic bonding of the benzene ring contribute to our calculated fluorescence rates. In the benzene ring, there are six such electrons that are delocalized and participate in the dominant transitions. Therefore, the number of target electrons here is $N_T^{\text{exc}} = 6 N_m$, where $N_m$ is the number of target molecules. For KamLAND, which contains PC and dodecane, we consider excitation only with PC, as dodecane is not benzene-based.

The particle physics model is encapsulated by the DM form factor $F_{\rm{DM}}(q)$, which contains the momentum dependence of the non-relativistic amplitude $\mathcal{M}_{e\chi}$ of DM scattering off free electrons~\cite{Essig:2011nj, Lee:2015qva},
\begin{equation}
    |F_{\rm{DM}}(q)|^2 = \overline{|\mathcal{M}_{e\chi}(q)|}\, / \, \overline{|\mathcal{M}_{e\chi}(\alpha m_e)|}.
\end{equation}
It is standard to define $\bar{\sigma}_e$ as 
\begin{equation}
    \bar{\sigma}_e \equiv \frac{\mu_{e\chi}^2}{16 \pi m_{\chi}^2m_e^2} \overline{|\mathcal{M}_{e\chi}(\alpha m_e)|^2}\,,
\end{equation}
where the bar indicates evaluation at $q=\alpha m_e$, where $\alpha$ is the fine structure constant, and $m_e$ is the electron mass. To remain model-independent, we consider generic DM form factors for heavy ($F_{\rm{DM}}(q)=1$) and light ($F_{\rm{DM}}(q)=(\alpha m_e)^2/q^2$) mediators. 

Finally, $\eta(v_{\text{min}})$ is the mean inverse velocity of the DM above a threshold velocity $v_{\text{min}}$, expressed as~\cite{Essig:2015cda}
\begin{equation}
\label{eq:eta}
    \eta(v_{\text{min}}) = \int{d^3v \ g_\chi(\vec{v})\frac{1}{v}\Theta(v-v_{\text{min}})},
\end{equation}
where $g_{\chi}(\vec{v})$ is the spherically-symmetric velocity distribution of DM in the halo. Typically, $g_{\chi}(\vec{v})$ is described by a Maxwell-Boltzmann distribution written as 
\begin{equation}
    g_{\chi}(\vec{v}) = \frac{1}{K}\exp\left(-\frac{|\vec{v}+\vec{v}_E|^2}{v_0^2}\right)\Theta(v_{esc}-|\vec{v}+\vec{v}_E|),
\end{equation}
where $v_{esc}=600 \ \rm{km}/s$ is the Milky Way escape velocity, $v_E = 240 \ \rm{km}/s$ is the Earth’s velocity in the Milky Way rest frame, and $v_0 = 230 \ \rm{km}/s$ is the local mean DM velocity. The normalization constant $K$ is determined by imposing $\int d^3v \  g_\chi(\vec{v}) = 1$. The minimum DM velocity $v_{\text{min}}^{ij}$ required for the DM to deposit an energy $E_d$ onto the electron is
\begin{equation}
    v_{\text{min}}^{ij} = \frac{E_d}{q}+\frac{q}{2 m_{\chi}},
\end{equation}
where $E_d = \Delta E_{ij}$ is the energy required for the transition $i \to j$ to occur.

To obtain the minimum and maximum values of $q$, we determine the values of $q$ for which $v_{\text{min}}^{ij}$ is equal to the maximum possible DM velocity, $v_E+v_{esc}$. When all of the DM’s kinetic energy is transferred to the electron, this sets the upper bound $E_d < \frac12 m_{\chi} (v_E + v_{\text{esc}})^2 \equiv E_d^{\text{max}}$. The minimum energy $E_d^{\text{min}}$ is the excitation threshold energy. Requiring $E_d^{\text{max}} > E_d^{\text{min}}$ sets a lower bound on $m_{\chi}$ for sensitivity to excitation, 
\begin{equation}
\label{eq:exc_min_mX}
    m_{\chi} \gtrsim \frac{2 \times 10^{-6} \ \text{MeV}}{(v_E+v_{\text{esc}})^2} \left(\frac{E_d^{\text{min}}}{\text{eV}}\right) \simeq 1.3 \ \text{MeV},
\end{equation}
where in the last equality we have specified $E_d^{\text{min}} \simeq 5 \ \text{eV}$, corresponding to the minimum excitation energy for the benzene-based targets we consider. It is worth noting that the specific transition energies and shape of the form factors for the organic scintillators used in these experiments will be somewhat different from our model. The simplified modeling on these materials is expected to dominate the uncertainty of the signal rate near the minimum visible DM mass.


\subsection{Ionization Rate}
\label{sec:diffrate}

The ionization rate for spin-independent DM-electron interactions is 
\begin{equation}
\label{eq:ionRate}
    R_{\rm{ion}} = \int \xi_{\rm{ion}}(E_{er}) \frac{dR_{\rm{ion}}}{d\ln E_{er}} \frac{1}{E_{er}}dE_{er},
\end{equation}
where $E_{er}$ is the electron’s recoil energy after ionization, $\frac{dR_{\rm{ion}}}{d\ln E_{er}}$ is the differential ionization scattering rate, and $\xi_{\rm{ion}}(E_{er})$ is the energy-dependent ionization efficiency factor. The differential scattering rate for DM-electron ionization is~\cite{Essig:2011nj, Lee:2015qva}
\begin{align}
\label{eq:diffRate}
    &\frac{dR_{\text{ion}}}{d\ln E_{er}} = \frac{\rho_{\chi} \bar{\sigma}_e}{8 m_{\chi} \mu_{e \chi}^2} F_{\rm Fermi}(k’) \\ &\times \sum_i N_T^i \int |f_{\rm{ion}}^{i}(E_{er},q)|^2 \ |F_{\text{DM}}(q)|^2 \ \eta(v_{\text{min}}(E_d)) \ q \ dq,\nonumber
\end{align}
where the differences from the excitation rate lie in the Fermi factor $F_{\rm Fermi}(k’)$ (discussed below), the number of targets $N_T^{i}$, the ionization form factor $f_{\rm{ion}}^{i}(E_{er},q)$, the deposited energy $E_d$, and the replacement of $\xi_{\text{exc}}$ with $\xi_{\rm{ion}}(E_{er})$ in Eq.~(\ref{eq:ionRate}). Moreover, instead of considering transitions between molecular orbitals, for ionization we focus on each atom individually. Prior to scattering, the electron is in a bound atomic state $i$. When the DM scatters off the electron, the electron may be ionized to a final state with momentum $k’$, assuming there is enough energy for ionization to occur. 

Considering that each state $i$ may have a different number of target electrons, we compute the total number of targets $N_T^i$ for ionization as 
\begin{equation}
    N_T^i = N_e^i \cdot N_a \cdot N_m,
\end{equation}
where $N_e^i$ is the number of electrons in state $i$, $N_a$ is the number of atoms in the target molecule, and $N_m$ is the number of target molecules in the detector (see Table~\ref{tab:detector_comparison}). For the targets we consider in Sec.~\ref{sec:detectors} (LAB, PC, dodecane), the bound states $i$ refer to the atomic orbitals of the carbon atom (1s, 2s, 2p) and the hydrogen atom (1s). 

The ionization form factor parametrizes the probability of absorbing momentum $q$, during the transition from an initial state $\Psi_i$ to an ionized state $\Psi_{\rm ion}$,
\begin{align}
\label{eq:ff_first}
    |f^i_{\rm{ion}}(E_{er},q)|^2 &\equiv  \delta(E_d - (E_{er} - E_b^i)) \nonumber \\
    &\times \left \vert \int d^{3}r \,e^{-i\mathbf{q}\cdot\mathbf{r}} \Psi_{\rm ion}^{*}(\mathbf{r})\Psi_{\rm i}(\mathbf{r}) \right \vert^2,
\end{align}
where the outgoing ionized wavefunction is approximated by a plane wave, $\Psi_{\rm ion} \propto \exp{i\mathbf{k'}\cdot\mathbf{r}}$. In Eq.~(\ref{eq:ff_first}), $E_d = E_{er} + E_{\rm{ion}}^i$ is the energy deposited by the DM, $E_b^i$ is the binding energy of the initial bound state $i$, and $E_{\rm ion}^i = - E_b^i$ is the corresponding ionization energy. As in Ref.~\cite{Blanco:2022cel}, we adopt a simplified model of molecules where the momentum is large enough to localize the interaction to a single atom, which occurs when $q\gg 2/a_0 \approx 7$ keV. Such is the case when the outgoing electron has keV-scale kinetic energy, which corresponds to the typical electron energy that produces a signal. With $E_{er} \sim \rm{keV}$, $q\approx \sqrt{2m_eE_{er}} \gtrsim 30$ keV. We can then approximate the initial state $\Psi_i$ with a simple atomic orbital, which takes the form of an electronic wavefunction $\Psi_{i,n \ell m}(\mathbf{r}) \equiv \mathcal{Y}_{\ell m}(\Omega)\chi_{n \ell}(r)$, where the radial function is
\begin{equation}
    \chi_{n\ell}(r) =\sum_j c_{j \ell n} N_{j\ell} r^{n_{j\ell}-1}\exp(-Z_{j\ell}r).  
\end{equation}
These are Roothaan-Hartree-Fock radial wave functions expanded as a sum of Slater-type functions, with quantum numbers $n$ and $\ell$, expansion coefficients $c_{j \ell n}$, normalization constants $N_{j\ell}$, and effective charges $Z_{j\ell}$. We obtain $c_{j \ell n}$, $N_{j\ell}$, and $Z_{j\ell}$ from the tables in Ref.~\cite{Bunge:1993jsz}. 

Because the final state of the electron after ionization can be approximated as a plane wave, its final momentum is $k'=\sqrt{2 m_e E_{er}}$. The ionization form factor is then~\cite{Lee:2015qva}
\begin{equation}
\label{eq:ff}
    |f_{\rm{ion}}^i(k',q)|^2 = \frac{(2\ell+1)k'^{2}}{4 \pi^3q} \int_{|k'-q|}^{k'+q} dk \ k|\tilde{\chi}_{n\ell}(k)|^2,
\end{equation}
where $\tilde{\chi}_{n \ell}(k)$ is the Fourier transform of the radial wave function~\cite{DževadBelkić_1989}. Note that the integral in Eq.~(\ref{eq:ff}) needs to be normalized such that $\int_{0}^{\infty} dk \ k^2|\tilde{\chi}_{n\ell}(k)|^2 = (2\pi)^3$. The form factor for each of the carbon orbitals can be calculated using Eq.~(\ref{eq:ff}); the hydrogen form factor is written in its closed form as per Ref.~\cite{Prabhu:2022dtm}. 

\begin{figure}[t]
    \centering
    \includegraphics[width=\columnwidth]{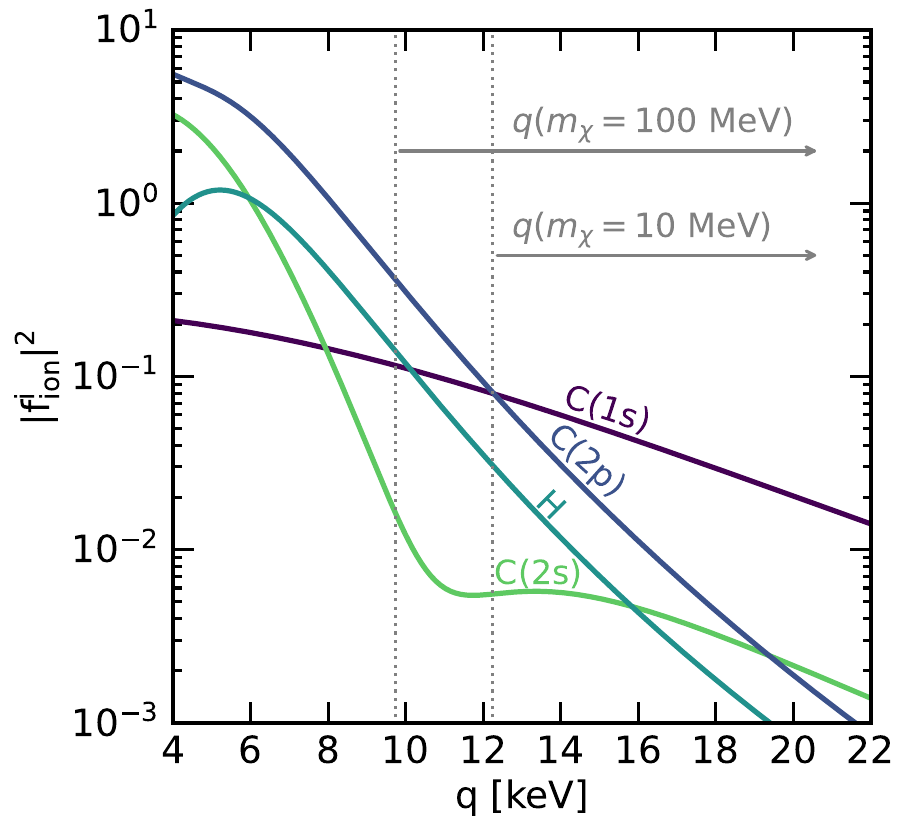}
    \caption{Ionization form factor for each orbital at recoil energy $E_{er} = 15$~eV. The dashed lines correspond to the minimum momentum transfer $q_{\rm{min}}$ for $m_{\chi}=10 \ \rm{MeV}$ and $m_{\chi}=100 \ \rm{MeV}$.}
    \label{fig:ff}
\end{figure}

\begin{figure*}[t]
    \centering
    \includegraphics[width=\textwidth]{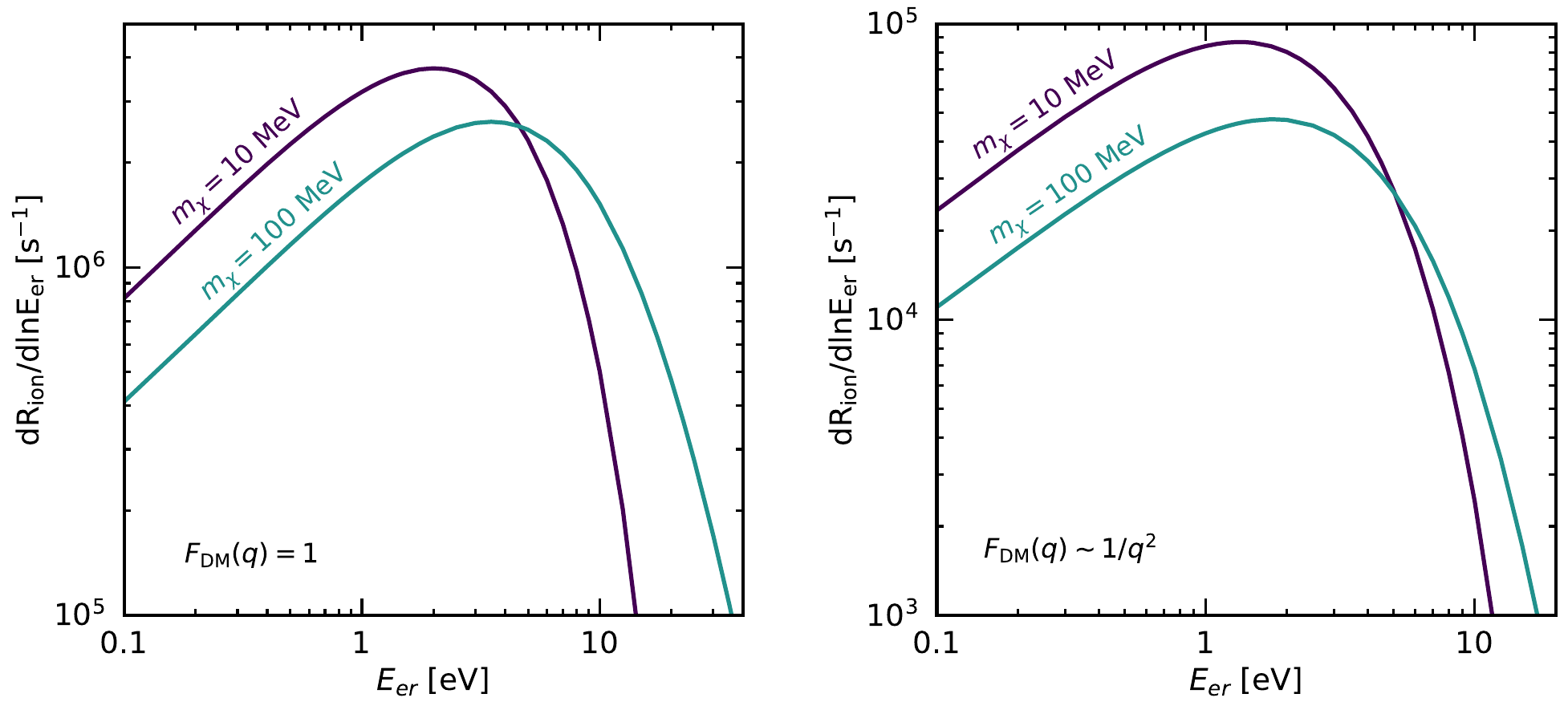}
    \caption{Differential event rate in JUNO from ionization by DM, as a function of the electron recoil energy, for form factors corresponding to a heavy mediator (left) and a light mediator (right). The differential rate is evaluated at $m_{\chi}=10 \ \rm{MeV}$ and $m_{\chi}=100 \ \rm{MeV}$ for a cross section of $\bar{\sigma}_{e}=10^{-35} \ \rm{cm}^2$. Note that there are different axes ranges between the two panels.}
    \label{fig:diffRate}
\end{figure*}

Figure~\ref{fig:ff} shows the ionization form factors for carbon orbitals and hydrogen. These form factors are evaluated at recoil energy $E_{er} = 15$~eV for easier comparison with Ref.~\cite{Lee:2015qva}. For reference, Fig.~\ref{fig:ff} also shows the minimum momentum transfer $q_{\rm{min}}$ needed for DM masses of $m_{\chi} = 10$~MeV and $m_{\chi} = 100$~MeV to ionize an electron in the weakest carbon bound state (2p) and have $E_{er} = 15$~eV of leftover energy. For heavier DM, more of the form factor's peak is included as a result of the greater energy imparted to the electron at a particular speed. As $E_{er}$ increases, both the form factor peak and $q_{\rm{min}}$ shift to slightly larger momentum transfers. 

The Fermi factor $F_{\rm Fermi}(k’)$ accounts for the wavefunction distortion of the ionized electron due to the presence of the nearby charged ion. The Fermi factor is
\begin{align}
\label{eq:FFermi}
    F_{\rm Fermi}(k') = \frac{\hat\Psi_{\rm ion}(\mathbf{r}=0)}{\Psi_{\rm ion}(\mathbf{r}=0)} 
    \approx\frac{2 \pi \nu}{1 - \text{exp}(-2 \pi \nu)},
\end{align}
where $\hat\Psi$ is the exact ionized wave function and $\nu = \alpha m_e Z_{\text{eff}}/k'$, where $Z_{\text{eff}}$ is the effective charge felt by the electron, which we set to $Z_{\text{eff}}=1$ to be conservative~\cite{Lee:2015qva, Blanco:2022cel}. 

In the same way that we determined the minimum DM mass required for excitation in Eq.~(\ref{eq:exc_min_mX}), we can also find the minimum DM mass for ionization by replacing the minimum deposited energy $E_d^{\min}$ with the lowest ionization energy of all atomic orbitals. Therefore, we now set $E_d^{\text{min}} \simeq 12 \ \text{eV}$, corresponding to the ionization energy of the 2p orbital in carbon. This leads to a minimum DM mass of about~$3 \ \text{MeV}$.

Following the arguments in Ref.~\cite{Essig:2015cda}, we can use fundamental kinematic principles to understand the relevant scales in DM-electron scattering. Because the electron's mean velocity ($\sim10^{-2}$) is roughly an order of magnitude larger than the DM's mean velocity ($\sim10^{-3}$), the relative velocity between the two particles is dominated by the electron’s velocity. Therefore, the scale of $q$ is set by the electron’s momentum, $p_e = m_e v_e \sim m_e \alpha \sim \text{keV}$. Another relevant scale to consider is that of the deposited energy $E_d$, which can be related to $q$ via energy conservation,
\begin{equation}
\label{eq:energyConserve}
    E_d = \vec{q} \cdot \vec{v} - \frac{q^2}{2 m_{\chi}}.
\end{equation} 
Assuming $v\sim10^{-3}$, $q\sim\text{keV}$, and $m_{\chi} \gg \ \text{MeV}$, the first term on the right in Eq.~(\ref{eq:energyConserve}) dominates and $E_d\sim qv \sim \text{eV}$. This implies that $E_d \gtrsim \text{eV}$ requires $v$ and $p_e$ to be on the upper ends of their respective distributions. 

Figure~\ref{fig:diffRate} shows the differential scattering rate given by Eq.~(\ref{eq:diffRate}) for different DM masses, in the heavy mediator case (left panel) and the light mediator case (right panel). While the differential rate includes a wider range of recoil energies $E_{er}$ for the heavier DM mass, the high $E_{er}$ have a negligible contribution to the total rate, as the peak occurs at low $E_{er}$. Thus, the lighter DM mass achieves a greater total rate due to its larger number density, producing a higher peak in the differential rate. 


\subsection{Light Yield and Detector Efficiency}
\label{sec:LY}

The efficiency factors $\xi_{\text{exc}}$ and $\xi_{\text{ion}}(E_{er})$ convert deposited energy into an expected detected signal by combining: (i) the \emph{intrinsic} scintillation/fluorescence yield of the medium, and (ii) the \emph{detector response} from photon transport through the scintillator to photoelectron (PE) production in the PMTs. We encapsulate the latter with the ratio $PE/LY$, where $LY$ is the intrinsic scintillation light yield (photons/keV) and $PE$ is the measured photoelectron yield (PE/keV), so that $PE/LY$ summarizes all optical and PMT effects (coverage, quantum efficiency, absorption/re-emission, attenuation,  etc.).

For excitation, the efficiency factor is
\begin{equation}
    \xi_{\text{exc}} = \frac{PE}{LY} Q,
\end{equation}
where $Q$ is the fluorescence quantum yield of the detector medium. We take $Q=0.2$ for LAB and $Q=0.4$ for PC~\cite{Buck:2015jxa}. These values are more accurate than the conservative generic choice $Q=0.1$ adopted in Ref.~\cite{Leane:2025efj}, and they account for the corresponding improvement of the excitation sensitivity relative to that work.

In practice, if the fluor concentration is sufficiently high, non-radiative energy transfer from the solvent to the fluor can enhance re-emission relative to the solvent quantum yield alone~\cite{Schoppmann:2022hst}. For example, in SNO+ the LAB$\rightarrow$PPO transfer efficiency was measured to be $78\%$ at $2$~g/L~\cite{SNO:2020fhu}; multiplying by the PPO quantum yield ($0.84$)~\cite{Buck:2015jxa} gives an effective re-emission probability of $\simeq 0.65$. Quantifying this effect for each detector would require \textit{in situ} measurements of transfer efficiencies and compositions, so for conservatism and uniformity we retain the medium-specific choices for $Q$ quoted above.

For ionization, the light yield is determined in a different way. The scintillation arises from the propagation of the ionized electron through the medium, producing photons at a rate set by the light yield. The expected detected signal therefore scales with the total number of scintillation photons produced, $LY \times E_{er}$, giving
\begin{equation}
\label{eq:effic}
    \xi_{\text{ion}}(E_{er})= \frac{PE}{LY} (LY \times E_{er}),
\end{equation}
i.e., $\xi_{\text{ion}}(E_{er})$ is the expected number of detected photoelectrons from an electron recoil of energy $E_{er}$. Liquid scintillators typically have $LY\sim \mathcal{O}(10)\,$photons/keV (Table~\ref{tab:detector_comparison})~\cite{Reina:2023sqc,SNO:2020fhu,Elisei:1997tw,Suzuki:2014woa,Inoue:2004wv}, with an overall uncertainty at the $\sim 10\%$ level~\cite{Li:2015phc,Elisei:1997tw,SNO:2020fhu}.

The measured photoelectron yields used in Table~\ref{tab:detector_comparison} are taken from Refs.~\cite{JUNO:2024fdc,JUNO:2025fpc,SNO:2025chx,DayaBay:2013yxg,Zhan:2015aha,DayaBay:2016ouy,Ghiano:2019cjy,BOREXINO:2023ygs,Suekane:2004ny,Inoue:2004wv,Slad:2016oqd}. Since these detectors typically achieve $PE \lesssim \mathcal{O}(1)\,$PE/keV, a recoil must deposit $\sim$keV-scale energy to produce an $\mathcal{O}(1)$-PE signal, motivating the keV-scale effective thresholds assumed in our detectability estimates.

Table~\ref{tab:detector_comparison} also lists the factors $X$ (PMT wall coverage) and $Y$ (spectrally averaged PMT response, including quantum efficiency and emission/absorption effects) as defined in Ref.~\cite{Leane:2025efj}. The product $X\times Y$ represents approximately the same end-to-end photon-to-PE acceptance captured by $PE/LY$. We find $PE/LY$ and $X\times Y$ agree within a factor of $\sim$$3$ across detectors, providing a useful consistency check. We nevertheless adopt $PE/LY$ in our baseline treatment because it is anchored directly to the measured PE yield and thus automatically incorporates detector-specific effects that may not be fully captured in a simplified $X\times Y$ decomposition. 

\begin{figure*}[t]
    \centering
    \includegraphics[width=\textwidth]{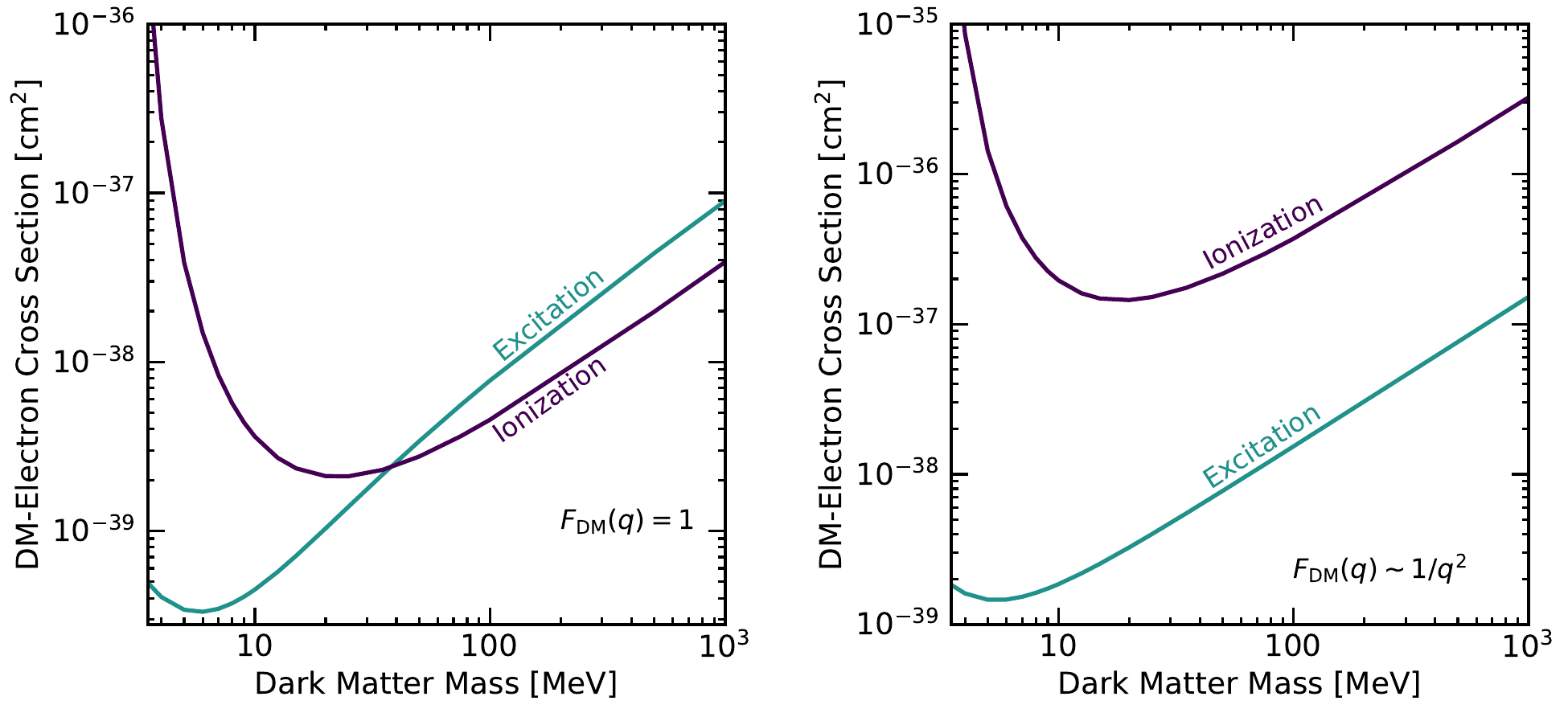}
    \caption{Sensitivity comparison between ionization or excitation alone in JUNO, for the heavy mediator (left) and light mediator (right). Note the different axis ranges between the two panels.}
    \label{fig:excVsIon}
\end{figure*}


\section{Dark Matter Sensitivity}
\label{sec:sensitivity}

As pointed out in Ref.~\cite{Leane:2025efj}, the annual modulation of the DM velocity relative to Earth can be used to handle the enormous dark rate of the PMTs in neutrino detectors relative to the DM signal rate. Because we are interested in the total number of PMT hits, the dark rate of the PMTs is our background. The annual modulation of the relative DM velocity has a distinct phase and peak time from most sources of background, making it possible to extract the DM signal. The relative DM velocity varies sinusoidally throughout the year by $15$ km/s, with the highest speed on June 2 and the lowest speed on December 2. The annual modulation is then 
\begin{equation}
    f_{\text{mod}} = \frac{R_{\text{high}}-R_{\text{low}}}{2\,R_0}\,,
\end{equation}
where $R_0$ is the rate assuming the standard $v_E = 240$ km/s, and $R_{\rm{high}(low)}$ is the rate obtained when adding (subtracting) $v_E$ by $15$ km/s. To set the 95\% confidence level (C.L.) sensitivity, we use the relation $S_{\text{tot}} = 2 \sqrt{B_{\text{tot}}}/f_{\text{mod}}$ as in Ref.~\cite{Leane:2025efj}, where $B_{\text{tot}}$ is the total background within the exposure time and $S_{\text{tot}}$ is the total rate of detectable DM scattering events (Eq.~(\ref{eq:totRate})) within the exposure time. 

It is important to note that the background rates we use in our simplified calculation here are estimated averages of the measured dark rates across different PMTs and times. In practice, the dark rates vary between PMTs and are not constant in time. In real experiments, one needs to perform various cleaning and detrending procedures (see Ref.~\cite{Leane:2025efj} for details); then a signal can be extracted using basic statistical methods (see, e.g., Ref.~\cite{Lisi:2004jw}). Multiple modulating backgrounds are also expected to be present, but these are sufficiently offset in time from the DM signal, and can also be measured with independent diagnostics; see the Supplemental Material of Ref.~\cite{Leane:2025efj} for more details.

We again emphasize that we are proposing to use the scalar PMT hit rate, which is the number of PMT hits per unit time~\cite{Leane:2025efj}. Note that this is distinct from the rate of PMT hits contributing to a triggered event, and requires negligible data storage. It is standard for neutrino detectors to record the scalar PMT rate to monitor individual PMTs. For example, this rate is publicly available for existing neutrino detectors including Borexino~\cite{Borexino:2013zhu}, Super-K~\cite{Super-Kamiokande:2023jbt}, and IceCube~\cite{IceCube:2011cwc}. It is also expected that JUNO will be able to record such data.  

Figure~\ref{fig:excVsIon} shows a comparison of the sensitivities from ionization and excitation in JUNO. In the heavy mediator case, excitation dominates at smaller DM masses due to the lower excitation threshold, and ionization dominates at larger DM masses as a result of the annual modulation (see Appendix). For the light mediator case, we find that including ionization has little effect on the sensitivity. In fact, the rates for ionization are much lower than for excitation. This is a result of the lower excitation threshold relative to ionization, leading to a wider range of momentum transfers in the excitation case. For the light mediator, the greater accessibility to lower momentum transfers from excitation has a significant effect on the sensitivity due to the $1/q^2$ scaling of the DM form factor. 

At first, one might expect that the ionization rate would dominate over the excitation rate at higher DM masses because ionization includes more possible final states than specific transitions between bound states. This would be true if increasing the DM mass continued to increase the amount of deposited energy at higher DM masses. However, as pointed out in Sec.~\ref{sec:diffrate}, the typical scale of the deposited energy is of order eV. Note that this fact is mostly independent of DM mass, given that the electron is the fastest particle in the problem and thus sets the energy scale. While it is possible for DM particles and electrons to move more quickly than their typical speeds and increase the deposited energy, these scenarios are not as likely. Therefore, more scattering events result in excitation than in ionization due to the lower excitation threshold. If there were only one electron per molecule, excitation would lead to a stronger sensitivity than ionization for both the heavy and light mediators. 

\begin{figure*}[t]
    \centering
    \includegraphics[width=\textwidth]{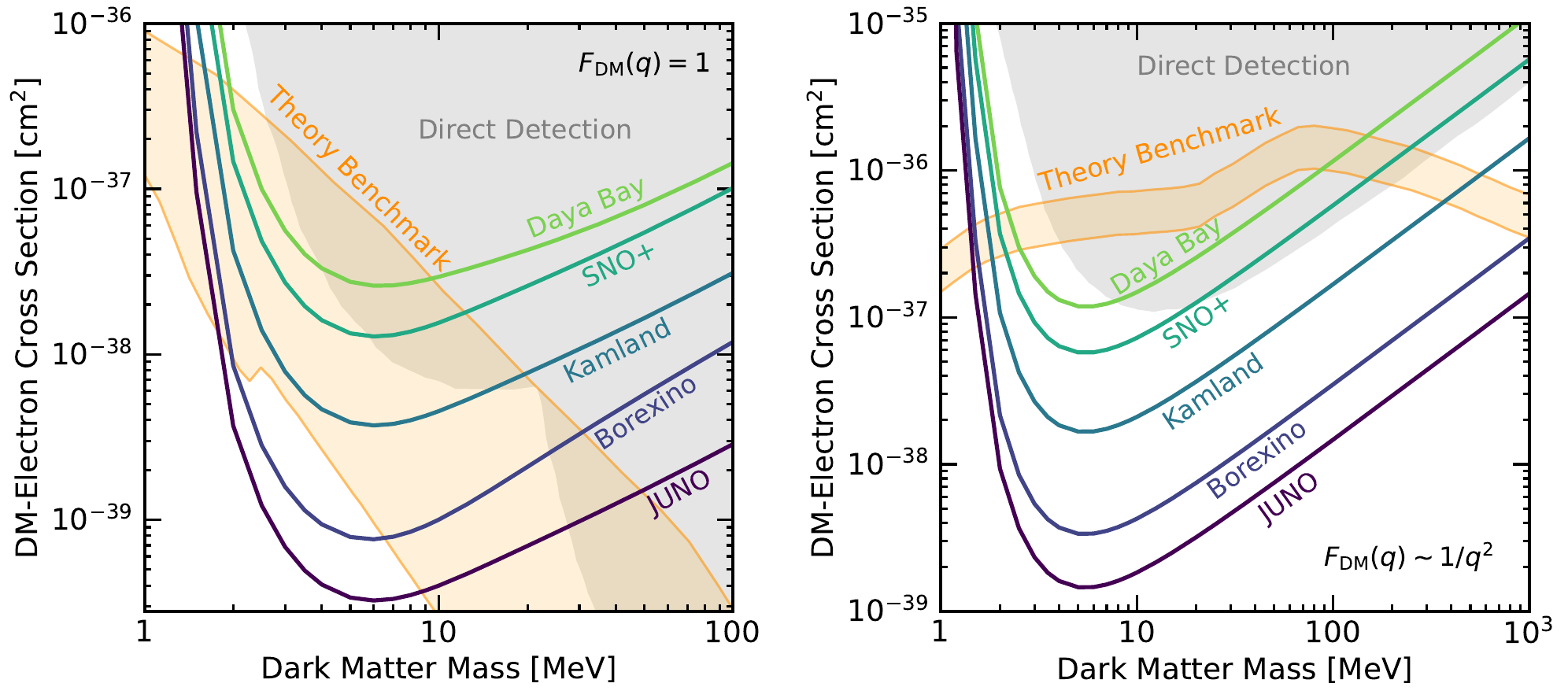}
    \caption{Potential sensitivity to DM-electron scattering cross sections at $95 \%$ C.L. for different neutrino detectors assuming an exposure of their runtime to date (except for JUNO, for which we assume 10 years of exposure). Sensitivities are shown for a heavy mediator (left) and a light mediator (right). In addition, direct detection limits~\cite{DAMIC:2019dcn, XENON:2019gfn, SuperCDMS:2020ymb, SENSEI:2020dpa, PandaX-II:2021nsg, DarkSide:2022knj, DAMIC-M:2025luv, Cheek:2025nul} are indicated by the gray shaded region. The orange shaded region labeled ``Theory Benchmark'' covers the parameter space corresponding to some particle-theory benchmark models~\cite{Essig:2022dfa}. Note the different axes ranges between the two panels.}
    \label{fig:sigmaTime}
\end{figure*}

Figure~\ref{fig:sigmaTime} shows the projected cross section sensitivity from both excitation and ionization resulting from DM-electron scattering for different neutrino detectors. The shape of the curve arises from a combination of the total event rate and the annual modulation. At lower DM masses, it becomes less likely that there will be sufficient energy for excitation or ionization, leading to a reduction in the event rate. At higher masses, the event rate approximately follows a $1/m_{\chi}$ scaling, reflecting the declining number density of DM particles. However, this scaling is not exact: for ionization, the light yield introduces some mass dependence. Additionally, the annual modulation becomes more pronounced at lower DM masses, as these masses rely more heavily on the high-velocity tail of the DM distribution to reach the excitation or ionization threshold. 

The shape of Borexino's curve in the heavy mediator case is slightly different from the others because excitation continues to dominate at higher masses for this detector, due to the fact that $Q$ is a factor of $2$ greater for PC than LAB. For KamLAND, which also has a PC component, the same effect does not occur because PC is only $20 \%$ of the target mass. Therefore, ionization dominates at higher masses for KamLAND in the heavy mediator case, as we also observe for the LAB-based detectors. In both the heavy and light mediator cases, the scaling differences between the detectors in Fig.~\ref{fig:sigmaTime} arise from several factors: target mass, background, light yield, efficiency, exposure time, and finally, when excitation dominates, quantum fluorescence yield.

For comparison, in Fig.~\ref{fig:sigmaTime} we also include direct detection constraints and theory benchmarks that produce the observed relic abundance. In the heavy mediator case, JUNO and Borexino can reach stronger sensitivities than direct detection by up to roughly an order of magnitude for DM masses less than 10 MeV. This is largely due to the higher excitation rates at lower DM masses. As the DM mass increases past 10 MeV, JUNO and Borexino continue to have stronger sensitivity until around 25 MeV, when the direct detection sensitivity takes over. For the light mediator, JUNO and Borexino exceed direct detection sensitivities by at least an order of magnitude for the entire mass range shown. 

Our results in Fig.~\ref{fig:sigmaTime} offer two new benefits compared to Ref.~\cite{Leane:2025efj}. One is that the contribution of both ionization and excitation provides a more complete picture of how the DM signal rate can be determined from the detected photons. The other is that the combination of multiple experiments leads to improved parameter inference. Because the experiments probe overlapping but distinct regions of parameter space, observing a signal in one experiment and not in another can help pinpoint specific DM masses and cross sections. With only an individual experiment, such as JUNO alone, one would only be able to infer that the signal lies above the minimum sensitivity curve, without clear information about the specific DM parameters. We also note that in Ref.~\cite{Leane:2025efj}, the adopted fluorescence quantum yield was $Q=0.1$, which is even more conservative than the more accurate values we use in this work ($Q$ = 0.2--0.4). In addition, Ref.~\cite{Leane:2025efj} calculated the efficiency of photon propagation and detection by considering $X \times Y$, while here we take the ratio $PE/LY$, as discussed in Sec.~\ref{sec:LY}. 


\section{Conclusions and Outlook}
\label{sec:conc}

Recently, there has been a surge of interest in sub-GeV DM, which is difficult to detect due to the low recoil energies falling below detector thresholds. New technologies have been developed to overcome this challenge, generally taking advantage of the more favorable kinematics in DM-electron scattering. 

To push to lower cross section sensitivities in this low-threshold regime, increasing the DM signal requires a larger target volume, which can be achieved by using neutrino detectors. In particular, neutrino observatories which are liquid scintillators also have the benefits of low excitation thresholds and high light yields. Ref.~\cite{Leane:2025efj} pointed out these advantages and showed how liquid scintillators can be used to search for sub-GeV DM, focusing on the excitation of LAB molecules in JUNO resulting from DM-electron scattering. 

In this work, we expand on the method developed in Ref.~\cite{Leane:2025efj}, considering several other neutrino detectors beyond JUNO and including contributions from both excitation and ionization in the DM signal rate. Although excitation has a lower threshold and leads to a higher light yield than ionization, the number of targets and annual modulation are higher for ionization, making it dominate over excitation at higher DM masses for the heavy mediator. In the case of the light mediator, excitation dominates over ionization due to the lower threshold and suppression at high momentum transfers. Furthermore, as we discussed in Sec.~\ref{sec:LY}, our excitation rates are conservative given our choice of the fluorescence quantum yield. 

We calculated projected sensitivities to the DM-electron cross section considering the runtime to date of the liquid scintillators SNO+, Daya Bay, Borexino, and KamLAND. For JUNO, we assumed a 10-year exposure. Our best sensitivities come from JUNO, because of its large target volume and high light yield and efficiency relative to the other detectors. At $m_{\chi} \simeq 6$ MeV, the sensitivity from JUNO reaches $\bar{\sigma}_{e} \simeq 3 \times 10^{-40}\, \rm{cm}^2$ for the heavy mediator. Borexino also has strong sensitivity at this mass, reaching $\bar{\sigma}_{e} \simeq 7 \times 10^{-40} \,\rm{cm}^2$. 

A coordinated multi-detector search is achievable with the liquid-scintillator detectors considered in this work. Notably, with multiple experiments having comparable sensitivity, they can cross-check each other to reduce systematic uncertainties and verify each other’s results. Because both the signal and background are influenced by the detector's properties, and the background is further impacted by the detector's location, it is important to have results from multiple experiments around the world. There is also an important opportunity for each of these experiments to develop and refine this new technique, removing the reliance on a single detector to implement it. With independent measurements in multiple detectors, a DM signal would not be limited to a single swath in the mass--cross section plane, and instead would produce multiple swaths that could intersect, aiding parameter inference.

Existing data from SNO+, Daya Bay, Borexino, and KamLAND make it possible for this analysis to begin already. In fact, with well over 10 years of exposure each, KamLAND and Borexino are particularly well-suited for this search. Multiple years of data will also be invaluable for understanding secular changes in the hit rate. All of this will get more exciting with JUNO having started this year. Together, these detectors offer a new path to reaching unprecedented sensitivity and robustness in sub-GeV DM searches.


\acknowledgements

We are grateful for helpful discussions with Mark Chen, Ben Lillard, and Bryce Littlejohn. LSO and RKL are supported by the U.S. Department of Energy under Contract DE-AC02-76SF00515. JFB is supported by National Science Foundation Grant No.\ PHY-2310018.


\clearpage
\onecolumngrid
\appendix
\section{Comparison between ionization and excitation rates}

\begin{figure*}[ht]
    \centering
    \includegraphics[width=17cm]{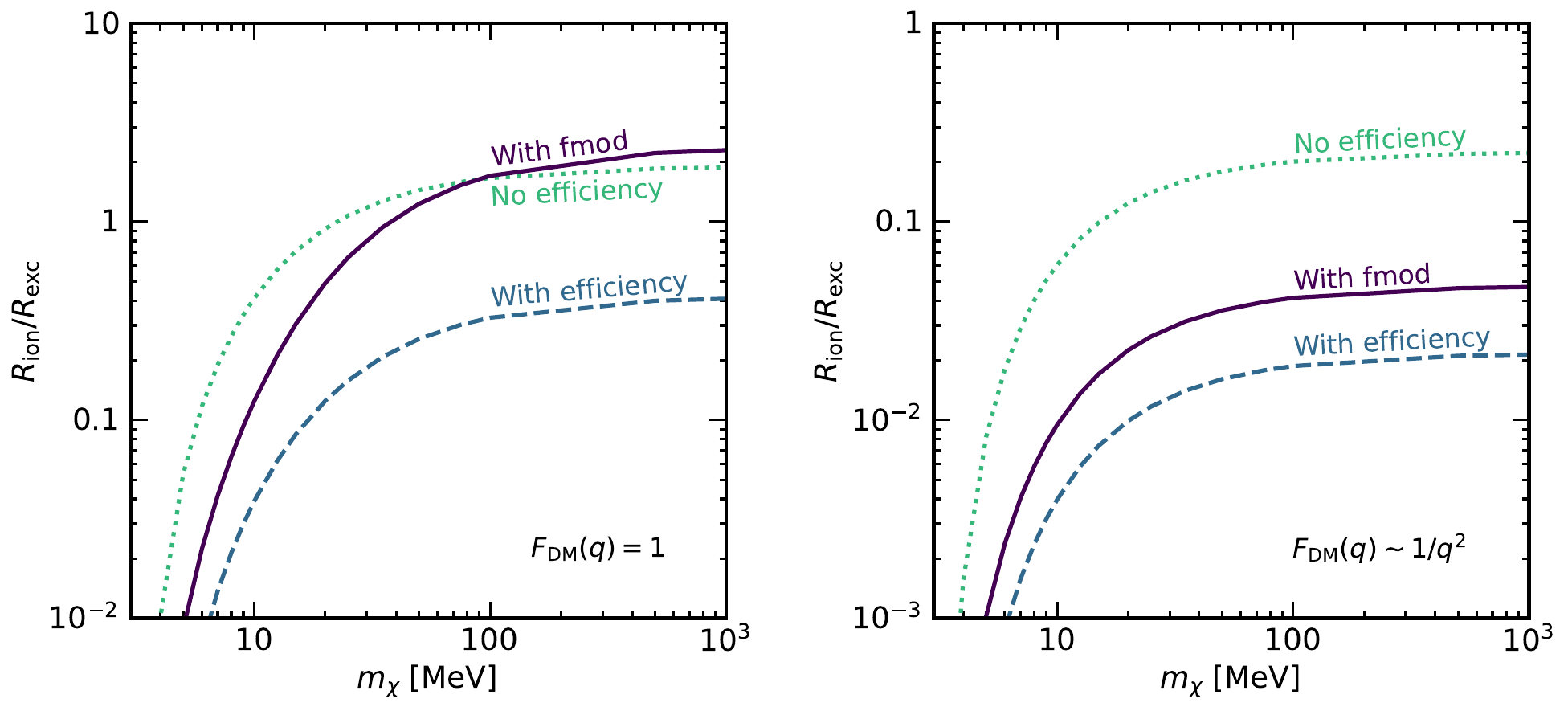}
    \caption{Ratio of $R_{\rm{ion}}$ to $R_{\rm{exc}}$ for JUNO with different factors included, for the heavy mediator (left) and light mediator (right). The dotted line is the ratio of the rates without the efficiency factors $\xi_{\rm{ion}}(E_{er})$ and $\xi_{\rm{exc}}$. The efficiency factors are included in the dashed line, which corresponds to the ratio of the rates given by Eq.~\eqref{eq:ionRate} and \eqref{eq:excRate}. Multiplying $R_{\rm{ion}}$ and $R_{\rm{exc}}$ by the corresponding $f_{\rm{mod}}$ (which encapsulates annual modulation) gives the ratio represented by the solid line. Note the different axis ranges between the two panels.}
    \label{fig:ratioIonExc}
\end{figure*}

Figure~\ref{fig:ratioIonExc} shows the ratio of the ionization rate to the excitation rate in JUNO under different conditions: without detector efficiency, with efficiency, and with annual modulation. When accounting for the larger number of target electrons contributing to the ionization rate (122 instead of 6) without incorporating light yield and efficiency, we find that for the heavy mediator, the ionization rate becomes slightly larger than the excitation rate at higher DM masses (see the dotted curve in the left panel of Fig.~\ref{fig:ratioIonExc}). However, including the light yield and efficiency factors brings the ratio back down to below one (see the dashed curve in the left panel of Fig.~\ref{fig:ratioIonExc}). This is because the quantum fluorescence yield is about an order of magnitude higher than the photon yield from ionized electrons at the relevant energies. Nonetheless, in the end ionization dominates the sensitivity for the heavy mediator at higher DM masses due to the annual modulation (see the solid curve in the left panel of Fig.~\ref{fig:ratioIonExc}). The annual modulation is higher for ionization because the threshold is higher, resulting in a greater dependence on the upper tail of the DM velocity distribution. 

\newpage


\bibliography{main}

\end{document}